\documentclass{emulateapj}
\usepackage{amsbsy}
\usepackage{amsmath}
\usepackage{graphicx}
\usepackage{amssymb}
\usepackage{rotating}
\begin{document}
\title{The Benefits of VLBI Astrometry to Pulsar Timing Array\\ Searches for Gravitational Radiation}
\author{
	D.R. Madison, S. Chatterjee, and J.M. Cordes\\
	Cornell University\\
	}
\begin{abstract}
Precision astrometry is an integral component of successful pulsar timing campaigns. Astrometric parameters are commonly derived by fitting them as parameters of a timing model to a series of pulse times of arrival (TOAs).  TOAs measured to microsecond precision over several-year spans can yield position measurements with sub-milliarcsecond precision. However, timing-based astrometry can become biased if a pulsar displays any red spin noise, which can be compared to the red noise signal produced by the stochastic gravitational wave background.  We investigate how noise of different spectral types is absorbed by timing models, leading to significant estimation biases in the astrometric parameters.  We find that commonly used techniques for fitting timing models in the presence of red noise (Cholesky whitening) prevent the absorption of noise into the timing model remarkably well if the time baseline of observations exceeds several years, but are inadequate for dealing with shorter pulsar data sets.  Independent of timing, pulsar-optimized very long baseline interferometry (VLBI) is capable of providing position estimates precise to the sub-milliarcsecond levels needed for high-precision timing.  In order to make VLBI astrometric parameters useful in pulsar timing models, the transformation between the International Celestial Reference Frame (ICRF) and the dynamical solar system ephemeris used for pulsar timing must be constrained to within a few microarcseconds.  We compute a transformation between the ICRF and pulsar timing frames and quantitatively discuss how the transformation will improve in coming years. We find that incorporating VLBI astrometry into the timing models of pulsars for which only a couple of years of timing data exist will lead to more realistic assessments of red spin noise and could enhance the amplitude of gravitational wave signatures in post-fit timing residuals by factors of 20 or more.  \end{abstract}
\maketitle

\section{Introduction}
Developing a mathematical model (a timing model) that predicts the time-varying, deterministic astrophysical effects that modulate pulse times of arrival (TOAs) is the practice of pulsar timing \citep{mt77,lk05}.  The main data products of a pulsar timing campaign are timing model parameters and a series of timing residuals---the differences between measured pulse arrival times and the predictions of a model.  A perfect timing model paired with a stable pulsar will yield timing residuals consistent with white noise.

Precisely known astrometric parameters (right ascension and declination $(\alpha,\delta)\equiv\boldsymbol{\theta}$, proper motion $(\mu_\alpha,\mu_\delta)\equiv$~$\boldsymbol{\mu}$, and parallax $\pi$) are essential to successful timing models and are useful in studies of the Galactic neutron star (NS) population \citep{cc97,tsb+99,fk07}, NS electromagnetic emission mechanisms \citep{dtb+09}, the Galactic distribution of interstellar gas \citep{cl02}, and the evolution of compact binaries containing pulsars, specifically to correct for the Shklovskii effect \citep{s69,dvt+08}.  A tenth of an arcsecond error in position can lead to annual oscillations in residuals of over 100~$\mu$s.  With software like TEMPO2 \citep{hem06}, timing models are refined through least-squares fitting to minimize the amplitude of such signatures in the residuals. For a particularly stable subset of pulsars, the millisecond pulsars (MSPs), measuring TOAs with microsecond precision over several years can yield position estimates with sub-milliarcsecond precision \citep{sns+05, hbo06,vbv+08}.

Pulsars are compact, point-like radio sources (smaller than a $\mu$as) that, if bright enough, allow precision astrometry with very long baseline interferometry (VLBI).  They are approximately a thousand times more compact than the active galactic nuclei (AGN) used as VLBI calibrators, so the precision of VLBI astrometry of pulsars is actually limited by evolution in the structure of AGN.  To enhance the S/N of VLBI observations of pulsars, gating (gathering signal only when the pulsar is beamed towards Earth) is used.  The Very Long Baseline Array (VLBA) allows for full-time VLBI and better-controlled systematics compared to other VLBI efforts owing to the identical antennas in the array.  With ever-advancing techniques and instruments like these, VLBI is capable of sub-milliarcsecond position estimates for some pulsars on par with the best timing-based results.  Additionally, VLBI astrometry can arrive at precisions requiring approximately five years of timing data in less than two years \citep[e.g.][]{cbv+09}. 

Canonical pulsars (CPs) and, to a lesser extent, MSPs display what is known as red noise in their timing residuals (it is called ``red'' because there is more power in low frequencies than in high frequencies).  For CPs, the red noise is from spin noise that may be caused by internal neutron star dynamics or changes in magnetospheric torque.  Past work has characterized spin noise with power-law spectra $P(f)\propto f^{-\gamma}$ with $4 \lesssim \gamma \lesssim 6$ for different pulsars.  More recently, discrete changes in spindown torque have been identified in some CPs \citep{klo+06,lhk+10} that in principle can be fitted for and removed.  In some cases, the torque appears quasi-periodic in a manner that is correlated with pulse-shape changes, suggesting the possibility that the torque variations can be removed in these cases as well.  If they cannot, discrete transitions in $\dot\nu$ collectively produce a red noise process with an $f^{-6}$ spectrum. \citet{sc10} reported a combined analysis of CPs and MSPs that included upper bounds on red noise in MSPs as well as measurements of significant red noise in two MSPs.  The combined analysis implies a global (multi-object) spectrum scaling as $f^{-5}$.  Red noise will produce stochastic shifts in the period or period derivative of a pulsar in a standard timing analysis by amounts that depend on how much power is absorbed into the sinusoidal astrometric fitting functions.  For short timing data sets (less than about 3 years), the absorbed power is large but becomes progressively smaller for longer data sets where the sinusoidal terms become progressively decoupled from the aperiodic spin noise.

Along with red spin noise, gravitational waves (GWs) induce structure in timing residuals.  There are current efforts to detect the signatures of isolated, continuous sources \citep{lwk+11,ejm12} and burst sources \citep{fl10,pbp10,vl10,cj12}, but among the most promising types of GW signals from an initial detection perspective is that of a stochastic background (SB) created by an ensemble of merging supermassive black holes (SMBHs) in the distant universe.  The SB is characterized by strains with an amplitude spectrum $h_c(f)\propto f^{-2/3}$ that manifest as a red process in timing residuals having a power spectrum $P(f)\propto f^{-13/3}$ \citep{jhv+06,svc08,cs12}. Because red spin noise and a SB of GWs both yield red timing residuals with similar spectral indices and possibly comparable magnitudes, it is exceedingly difficult to make a detection of this type of GW in the residuals of a single pulsar.  Fortunately, \citet{hd83} predict that a specific pattern of correlations unique to GWs can be observed in the residuals of an array of pulsars in the presence of a SB of GWs.  The European Pulsar Timing Array \citep[EPTA;][]{vlj+11}, the Parkes Pulsar Timing Array \citep[PPTA;][]{ych+11,mhb+13}, and the North American Nanohertz Observatory for Gravitational Waves \citep[NANOGrav;][]{dfg+13} are timing arrays of pulsars and looking for, among other things, the type of correlations in their residuals predicted by Hellings and Downs.

Whatever structure in timing residuals one is hoping to study, the model fitting applied may undermine those efforts.  Every parameter that is fitted reduces the degrees of freedom in the data set.  Furthermore, if one produces an initial timing model with approximately correct parameters, the timing residuals will contain astrophysical information, but will also contain the signatures of errors in the timing model.  Errors in each of the timing model parameters leave characteristic signatures in the timing residuals that have associated structures in frequency space.  Minimizing the amplitude of the signature of a timing model parameter error indiscriminately removes power from the residuals at the frequencies associated with that particular signature \citep[e.g.][]{brn84}.  The power that is removed from the pre-fit residuals is absorbed by the timing model and causes the parameters of the timing model to shift, possibly away from their true values \citep{emv11}.  \citet{chc+11} describe a generalized least-squares procedure (often referred to as Cholesky whitening) for conducting timing model fits in the presence of correlated red timing noise.  These techniques, by accounting for correlated structure in the noise, reduce the scale of noise-induced bias in parameter estimation and provide more realistic error estimates, often by yielding larger error estimates.  Though Cholesky techniques are an improvement over standard least-squares fitting techniques, properly modeling the covariance of the noise can prove tricky, and they are by no means a panacea for the problems introduced by red noise. 

In this paper, we investigate the biases induced in the astrometric parameters of a timing model when they are fit in the presence of some red process.  In so doing, we demonstrate how red processes can be partially absorbed into a timing model and thus attenuated in the residuals.  We do this because of the inherently red spectra of the spin noise associated with many pulsars and the SB of GWs, but also because the precision astrometric parameters needed to do high-quality pulsar timing can now be provided by VLBI.  Some of the deleterious effects of standard model fitting can be side-stepped if VLBI astrometry can be incorporated into pulsar timing campaigns.

In Section 2, we discuss astrometry as it pertains to pulsar timing models, and in particular, derive the signatures in timing residuals associated with incorrect estimates of position, proper motion, and parallax.  In Section 3, we describe simulations we conducted to assess the effects that fitting in the presence of red processes can have on the astrometric parameters of a timing model and on the signature of the red process itself.  We follow this with a discussion of our results in Section 4.  In Section 5, we discuss the real-world feasibility and benefits of incorporating VLBI astrometry into pulsar timing campaigns.  In Section 6, we discuss several applications for VLBI astrometry in pulsar science that are not directly tied to gravitational wave detection.  Finally, in Section~7, we summarize and provide some concluding remarks.

\section{Astrometric Terms in Timing Models}

In this section, we discuss the roles of position, proper motion, and parallax in the development of timing models.  For a more thorough description of this material, see \citet{bh86} and \citet{ehm06}. The position and proper motion principally affect the Roemer delay. The Roemer delay is the difference between the arrival time of a pulse at an observatory and the arrival time of that pulse at the solar system barycenter (SSB) from geometric path length difference.  It can be written as $\Delta_R =-\boldsymbol{r}\boldsymbol{\cdot}\boldsymbol{\hat{R}_{P}}/c$ where $\boldsymbol{r}$ is the vector pointing from the SSB to the geocenter (we consider errors in the topocentric to geocentric transformation to be negligible), and $\boldsymbol{\hat{R}_{P}}$ is the unit vector pointing from the observatory to the pulsar at the time of the observation.  

We use equatorial coordinates $\alpha$ and $\delta$ (referenced to the geocenter) in all other sections of this paper, but here, we use an ecliptic coordinate system (centered at the SSB) with longitude $\lambda$ and a latitude $\beta$.  Let $\lambda$ increase with the orbital motion of Earth from an $x$-axis that points from the Earth to the SSB at the vernal equinox.  Orient the $z$-axis such that the Earth orbits in a right-handed fashion.  Exploiting the relation $\boldsymbol{r}+\boldsymbol{R_P}=\boldsymbol{r_P}$, where $\boldsymbol{r_P}$ is the vector pointing from the SSB to the pulsar, and approximating the orbit of the Earth as circular, we write
\begin{eqnarray}
\boldsymbol{r}&=&-\hat{x}R\cos{(\omega t)}-\hat{y}R\sin{(\omega t)},\\
\boldsymbol{r_P}&=&\hat{x}D\cos{\beta}\cos{\lambda}+\hat{y}D\cos{\beta}\sin{\lambda}+\hat{z}D\sin{\beta},\\
\boldsymbol{R_P}&=&\hat{x}\left[D\cos{\beta}\cos{\lambda}+R\cos{(\omega t)}\right]+\nonumber\\&&\hat{y}\left[D\cos{\beta}\sin{\lambda}+R\sin{(\omega t)}\right]+\hat{z}D\sin{\beta},
\end{eqnarray}
where $R$ is 1 A.U., $\omega$ is $2\pi$ yr$^{-1}$, and $D$ is the distance from the SSB to the pulsar. If the position and proper motion of the pulsar are determined at $t=0$ and if any acceleration of the pulsar (from motion in a binary, Galactic rotation, etc.) is ignored, the position can be written as $\beta(t)=\beta_0+\mu_\beta t$ and $\lambda(t)=\lambda_0+\mu_\lambda t$ where $\mu_\lambda$ and $\mu_\beta$ are the components of the proper motion.  By expanding to linear order in the small parameters $\mu_\lambda, \mu_\beta$, and $\pi\equiv R/D$ and dropping any constant terms, the Roemer delay can be written as
\begin{eqnarray}
\Delta_R&\approx&\frac{R}{c}\left[\cos{\beta_0}\cos{(\omega t-\lambda_0)}\right.\nonumber\\
&&~~ -\mu_\beta t\sin{\beta_0}\cos{(\omega t-\lambda_0)}\nonumber\\
&&~~ -\mu_\lambda t\cos{\beta_0}\sin{(\omega t-\lambda_0)}\nonumber\\
&&~~ \left.-\frac{\pi}{2}\cos^2{\beta_0}\cos{(2\omega t-2\lambda_0)} \right].
\end{eqnarray}

Small offsets in the astrometric parameter estimates from their true values lead to incorrect modeling of the Roemer delay and produce signals in timing residuals that are sinusoidal with periods of one year and six months (position and parallax, respectively) and a linearly growing sinusoid with a one-year period (proper motion).  This is all demonstrated in Figure 1.  If a set of residuals display structure resembling a superposition of these signatures, fitting the model to the residuals will lead to adjustments of the astrometric parameters to flatten the residuals and make them more consistent with white noise.

\begin{figure}[tbp]
\begin{center}
\includegraphics[height = 70mm, width = 87mm]{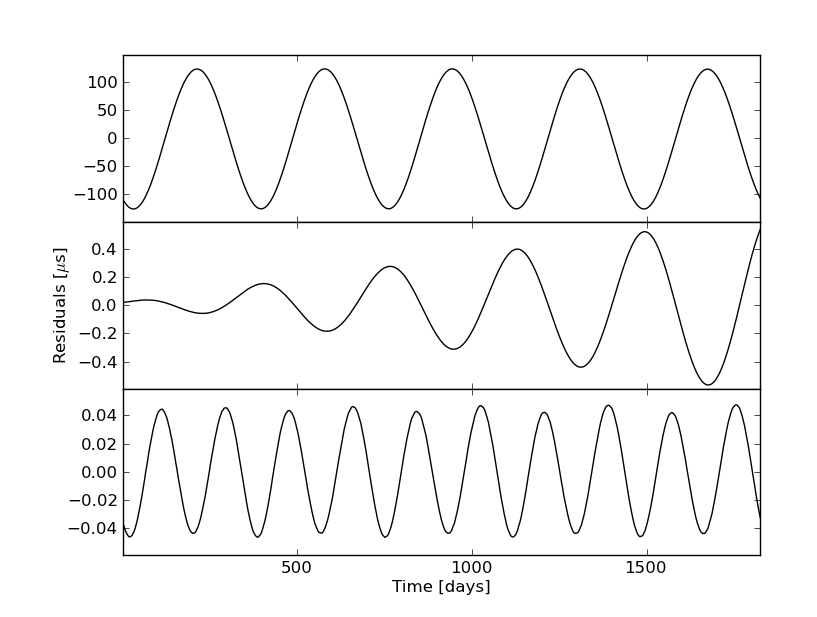}
\caption{The characteristic signatures in timing residuals caused by incorrect estimates of position, proper motion, and parallax (from top to bottom respectively). For a pulsar with the coordinates of PSR~J1713+0747, these are the residuals associated with 100 mas offsets in $\delta$ and $\pi$ and 100 mas yr$^{-1}$ offsets in $\mu_\delta$.  The oscillations in the top two plots have a 1 year period and the oscillations in the bottom plot have a 6 month period.  The amplitude of these curves scales linearly with the magnitude of the astrometric parameter bias so long as the error is small.}
\end{center}
\end{figure}

\section{Simulations of Noise-Induced Astrometric Biases \& Red Noise Absorption} 
We simulated three sets of noiseless arrival times for a pulsar with the astrometric parameters, spin frequency, and spin-down rate of PSR J1713+0747.  We did this with the ``fake'' plug-in for TEMPO2 \citep{hem06}. In the first set of pulse arrival times (Set A) we simulated one TOA every 3 days over a 500 day span while in the second set (Set B) we simulated one TOA every 21 days over a 500 day span.  In the final set (Set C) we simulated one TOA every 21 days over a 3,000 day span.  

We then added noise of different amplitudes and spectral characteristics to the simulated TOAs.  We considered red noise with a power spectrum $P(f)\propto f^{-\gamma}$ with spectral indices $\gamma=$ 0, 3, and 5.  We generated realizations of red noise in the Fourier domain by generating white noise-like spectra, multiplying them by the desired power-law shape, carrying out an inverse Fourier transform, and scaling the noise to produce our desired RMS.  We enforce that the realizations of red noise have zero mean.  With this technique for noise generation, we lose some very low-frequency power between zero frequency and our lowest frequency bin; this power will manifest itself as a nearly linear trend in the data that will be removed by fitting for the spin parameters of the pulsar.  For each $\gamma$, we chose 16 amplitudes for the pre-fit RMS of the noise logarithmically spaced between 10 ns and 10~$\mu$s.  Once combined with a particular realization of noise of a specific spectral type and RMS amplitude, we reprocessed the simulated pulse arrival times with TEMPO2, fitting for the phase ($\phi$), pulse period ($P$), pulse period derivative ($\dot{P}$), and astrometric parameters.  We recorded the RMS of the post-fit residuals along with the post-fit astrometric parameters.  For each spectral type of noise and each pre-fit noise amplitude, we did this with 100 different realizations of noise. 

Whenever we conducted simulations with correlated red noise, we carried out both standard weighted least-squares fits and generalized least-squares fits as described in \citet{chc+11}.  The key step to the generalized least-squares technique is to estimate the power spectrum of the noise in the residuals; we were aided in this as we controlled the power spectrum of the noise.  The Fourier transform of the power spectrum provides the covariance matrix, $\boldsymbol{\rm C}$, of the residuals.  Once $\boldsymbol{\rm C}$ is estimated, it can be decomposed as $\boldsymbol{\rm C} = \boldsymbol{\rm U}\boldsymbol{\rm U^T}$ through a Cholesky decomposition.  To check that the spectral estimation of the noise covariance matrix has not been biased by spectral leakage, the Cholesky matrix $\boldsymbol{\rm U}$ can be applied to the post-fit residuals $\boldsymbol{\rm E}$ to form $\boldsymbol{E_W} = \boldsymbol{\rm U^{-1}}\boldsymbol{\rm E}$.  If the covariance matrix has been appropriately modeled, the Cholesky-whitened residuals $\boldsymbol{E_W}$ should be consistent with white noise.  We found that to consistently and successfully whiten the residuals, they had to initially contain some white noise power.  For this reason, each residual we simulate has added to it a zero-mean Gaussian error with a 1 ns RMS. 

We calculated the post-fit biases in position, proper motion, and parallax as 
\begin{eqnarray}
\delta\theta &=&\left[(\alpha-\alpha_{post})^2\cos^2{\delta}+(\delta-\delta_{post})^2\right]^{1/2},\\
\delta\mu &=& \left[(\mu_{\alpha}-\mu_{\alpha,post})^2\cos^2{\delta}+(\mu_{\delta}-\mu_{\delta,post})^2\right]^{1/2},\\
\delta\pi &=& |\pi-\pi_{post}|.
\end{eqnarray}
Here, $\alpha$, $\delta$, $\mu_\alpha$, $\mu_\delta$, and $\pi$ are the astrometric parameters used to simulate the initial TOAs; the quantities with a subscript ``$post$'' are the post-fit estimates for them.  In a similar fashion, we calculated the estimated astrometric parameter uncertainty based on the parameter uncertainties from the fit.  For example, $\sigma_\theta = [\sigma_\alpha^2\cos^2{\delta}+\sigma_\delta^2]^{1/2}$ where $\sigma_\alpha$ and $\sigma_\delta$ are the 1-$\sigma$ uncertainties on $\alpha$ and $\delta$ provided by TEMPO2.

\begin{figure*}[h!tbp]
\begin{center}
\includegraphics[height = 68mm, width = 86mm]{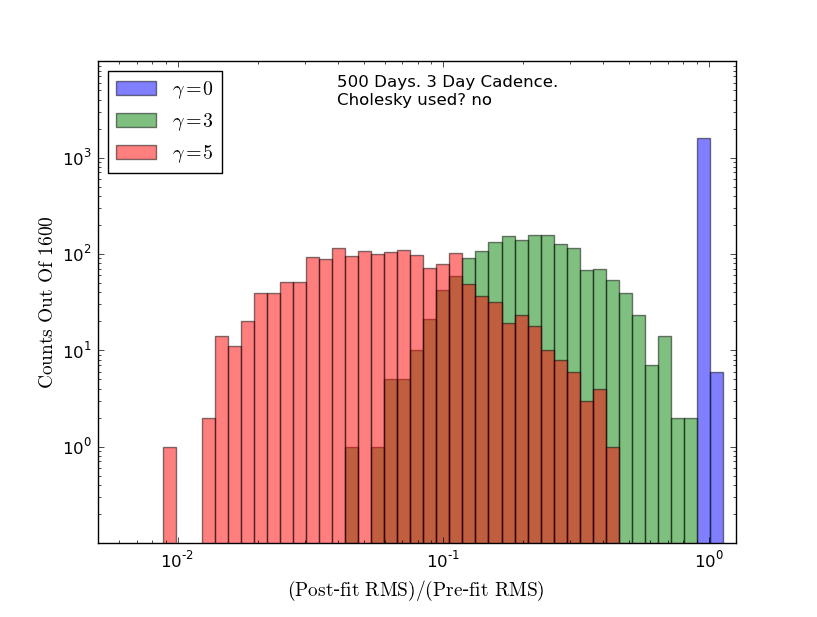}
\includegraphics[height = 68mm, width = 86mm]{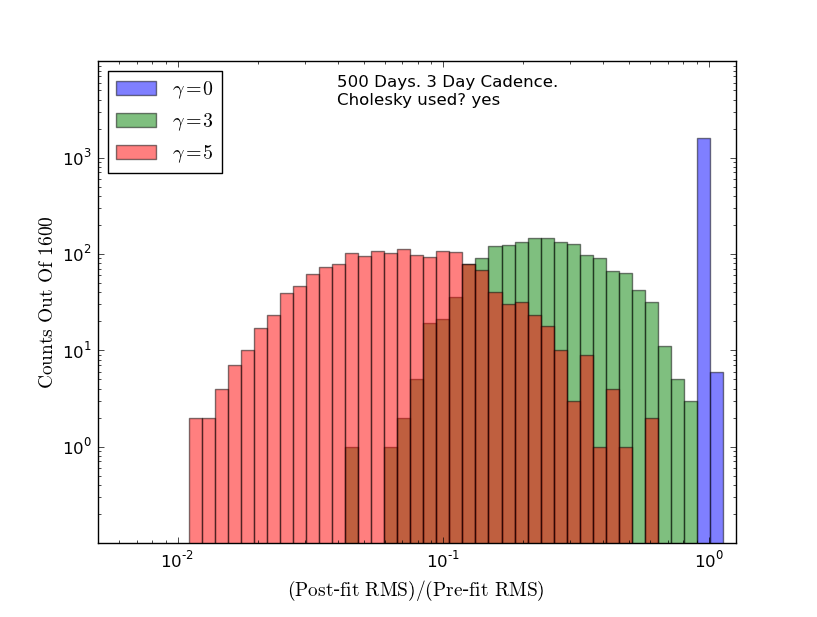}\\
\includegraphics[height = 68mm, width = 86mm]{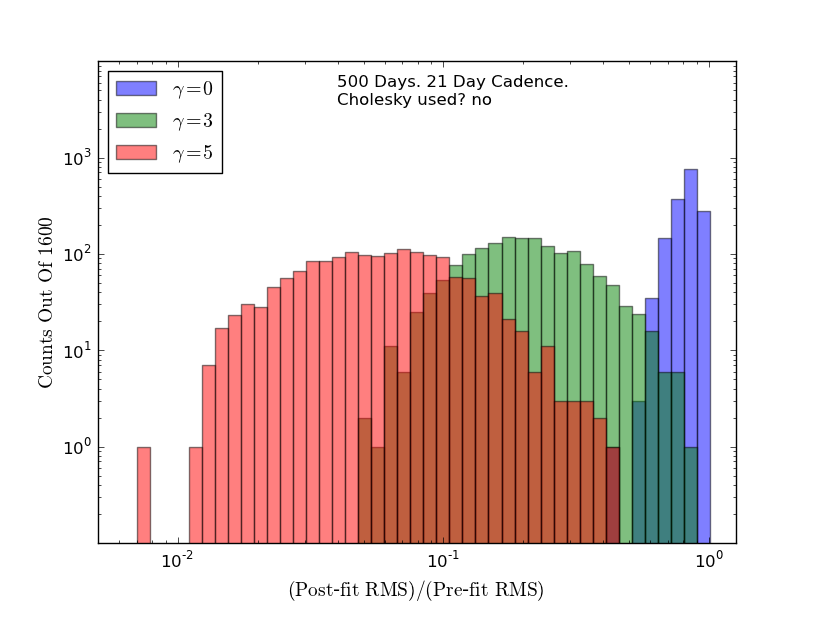}
\includegraphics[height = 68mm, width = 86mm]{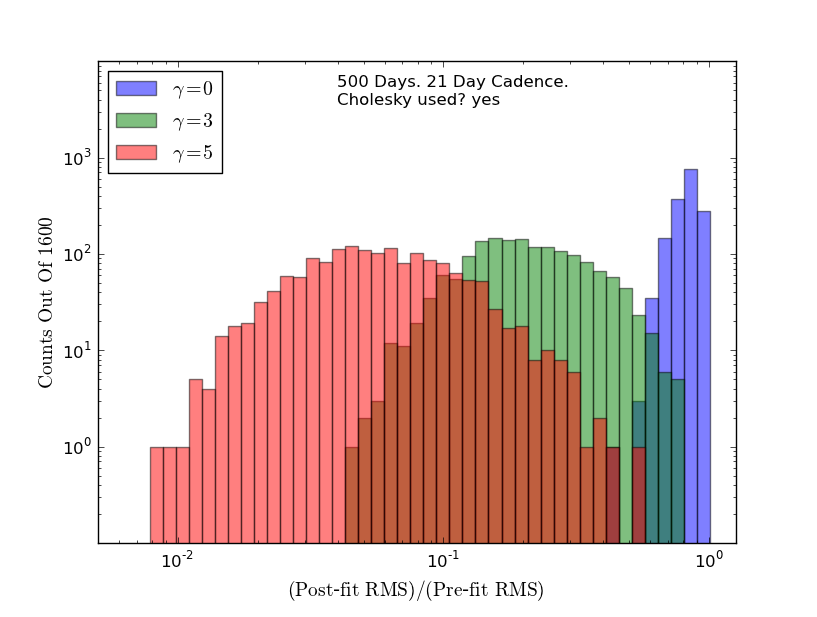}\\
\includegraphics[height = 68mm, width = 86mm]{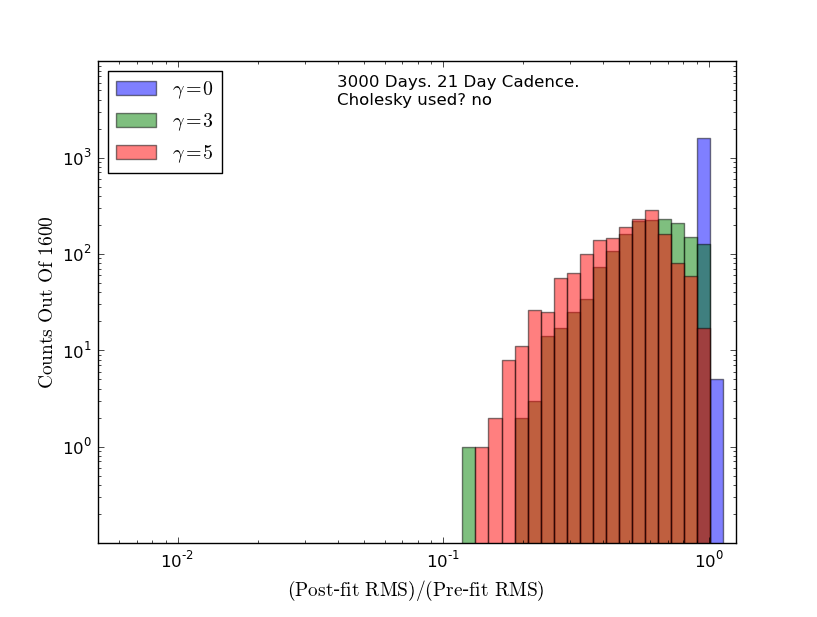}
\includegraphics[height = 68mm, width = 86mm]{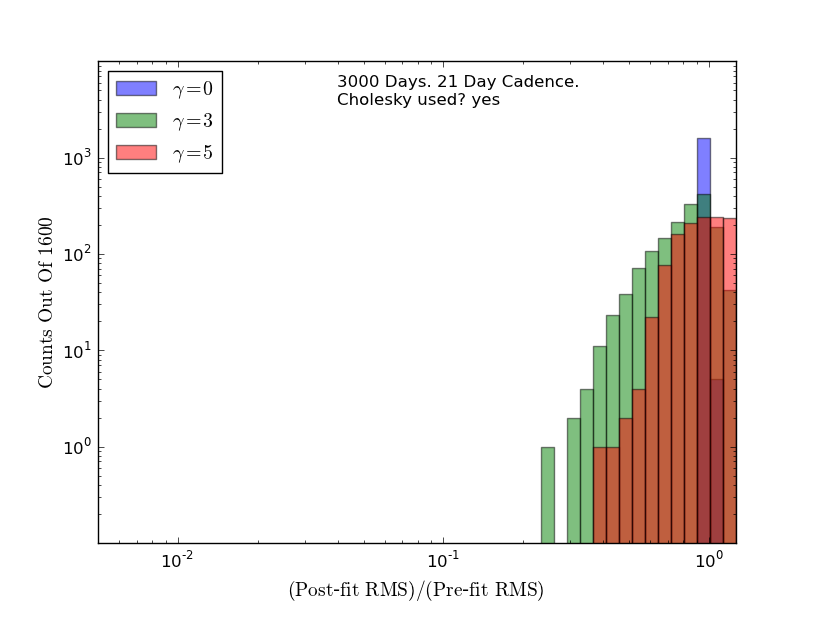}
\caption{Ratios of post-fit residual RMS to pre-fit residual RMS for simulated arrival times.  Each of the three histograms in each plot corresponds to input noise of a different spectral index ($\gamma=0,3,5$).  Each histogram contains 1600 counts for the 100 realizations of noise at each of 16 pre-fit RMS amplitudes.  The plots on the left were derived from simulations using standard linear least-squares fitting.  The plots on the right were derived from simulations using generalized least-squares fitting (Cholesky whitening).}
\end{center}
\end{figure*}

\begin{figure*}[h!tbp]
\begin{center}
\includegraphics[height = 68mm,width = 86mm]{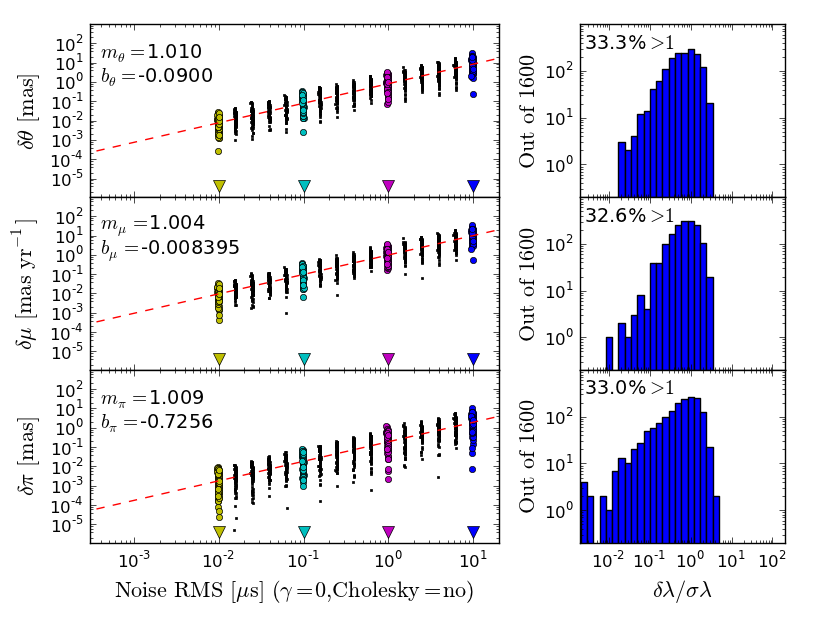}\\
\includegraphics[height = 68mm,width = 86mm]{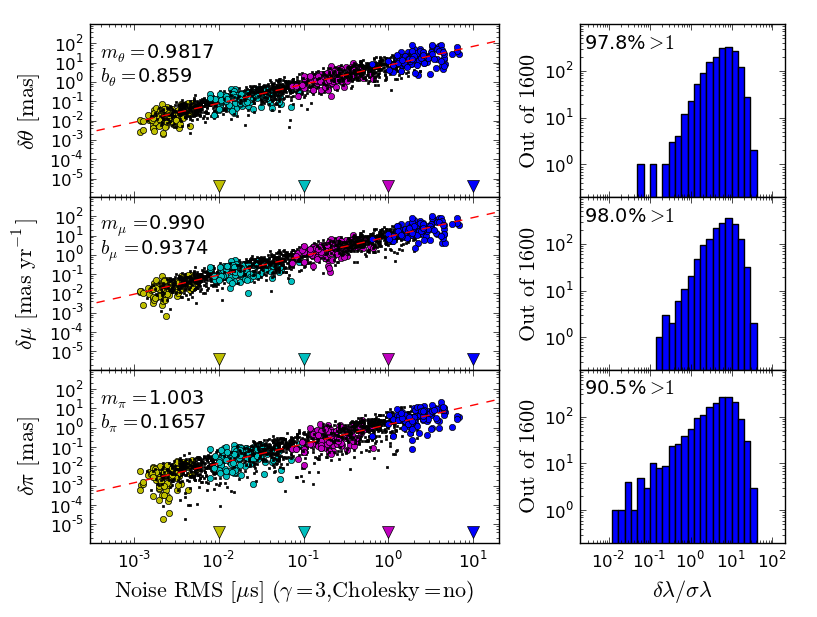}
\includegraphics[height = 68mm,width = 86mm]{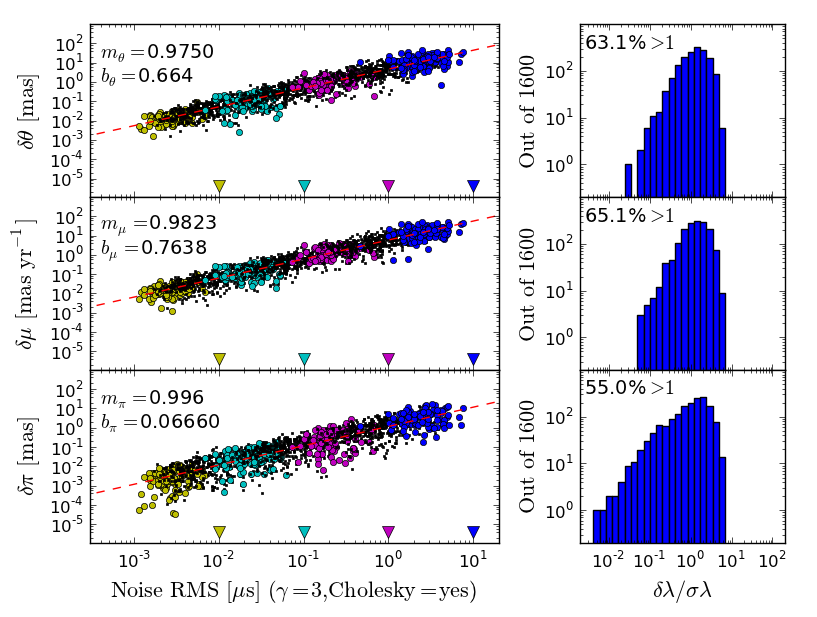}\\
\includegraphics[height = 68mm,width = 86mm]{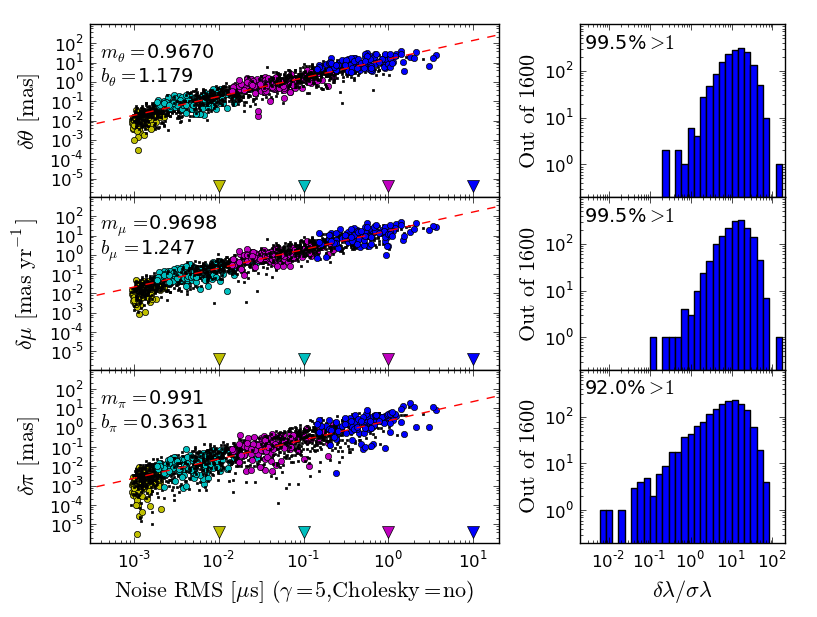}
\includegraphics[height = 68mm,width = 86mm]{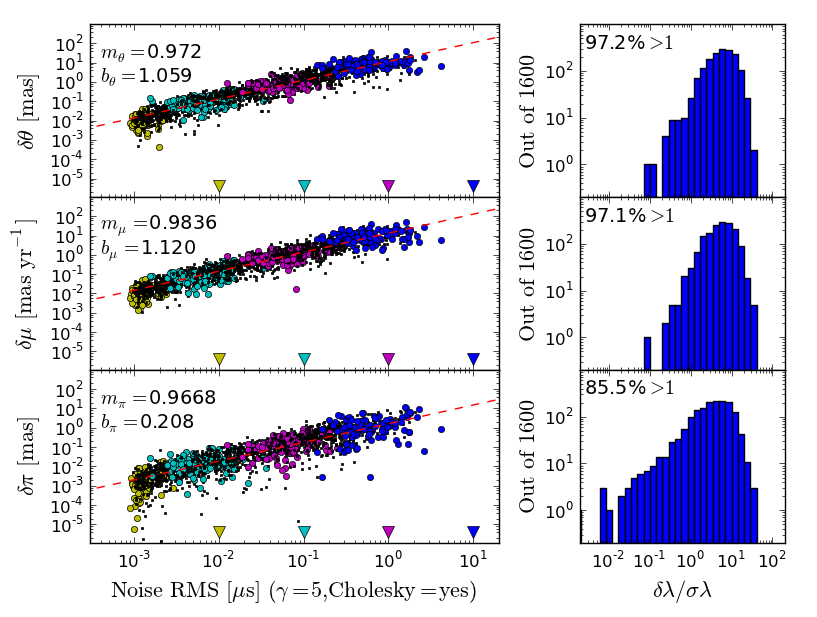}
\caption{Offsets between true astrometric parameters and post-fit estimates of astrometric parameters versus post-fit RMS for simulation Set A (500 days of TOAs taken at 3 day intervals).  The left three panels in each of the five plots each contains 1600 points corresponding to the 100 realizations of noise at 16 input amplitudes of pre-fit noise.  The yellow points correspond to an input pre-fit RMS of 0.01 $\mu$s (as indicated by the triangular yellow indicator above the horizontal axis).  Similarly, cyan, magenta, and blue points correspond to input pre-fit RMS values of 0.1, 1, and 10 $\mu$s respectively.  The dashed red lines are the least-squares best-fit scaling relations as described in Eq.~8 (the best-fit scaling parameters are listed above and to the left of the line).  The top, middle, and bottom left panel of each plot displays the noise-induced post-fit offset in the position $\theta$, proper motion $\mu$, and parallax $\pi$ respectively as described in Eqs. 5-7.  To the right of each scatter plot is a histogram of the offsets in that post-fit astrometric parameter ($\delta\lambda$) divided by the formal uncertainty in that parameter ($\sigma_\lambda$).  The percentage in the top left corner of each histogram is the percentage of the 1600 points in which this ratio exceeded 1.  In the top plot, only white noise was used ($\gamma=0$).  Moving down through the two lower rows of plots, the spectral index $\gamma$ of the simulated pre-fit residuals increases to 3 and then to 5.  In the plots in the left column, Cholesky whitening was not used.  In the plots in the right column, Cholesky whitening was used.}
\end{center}
\end{figure*}

\begin{figure*}[h!tbp]
\begin{center}
\includegraphics[height = 74mm,width = 86mm]{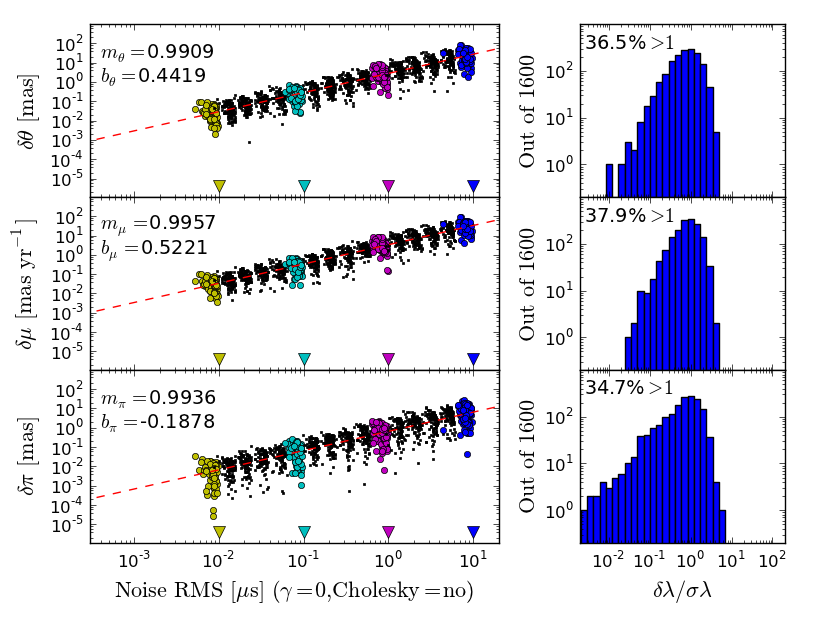}\\
\includegraphics[height = 74mm,width = 86mm]{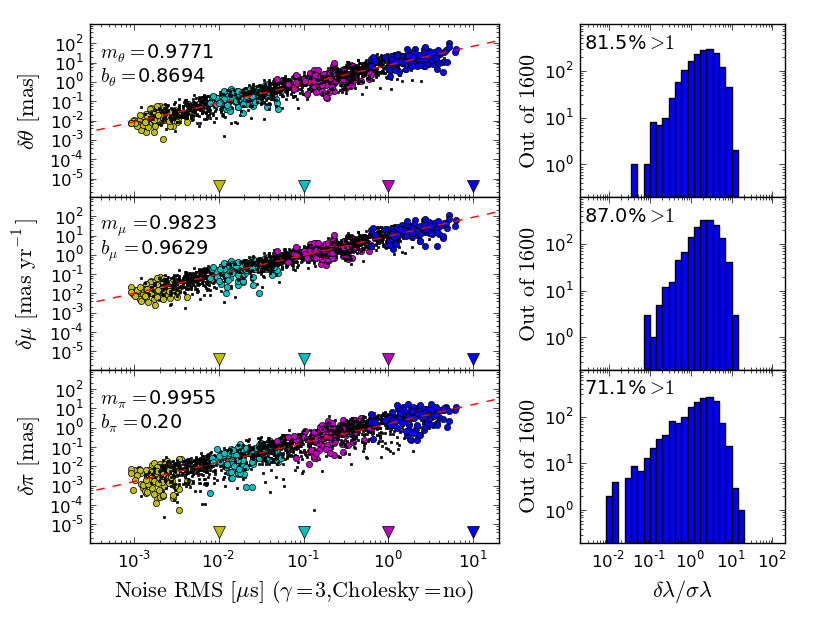}
\includegraphics[height = 74mm,width = 86mm]{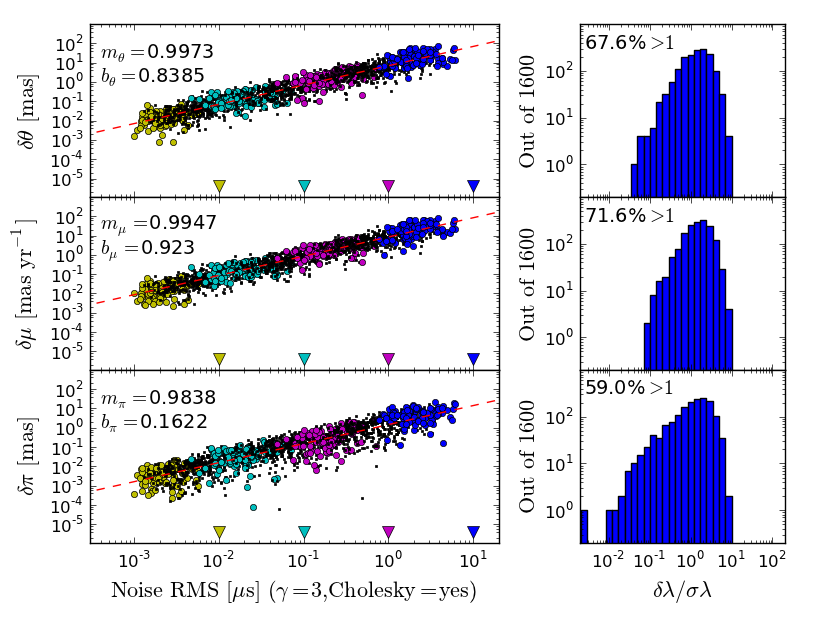}\\
\includegraphics[height = 74mm,width = 86mm]{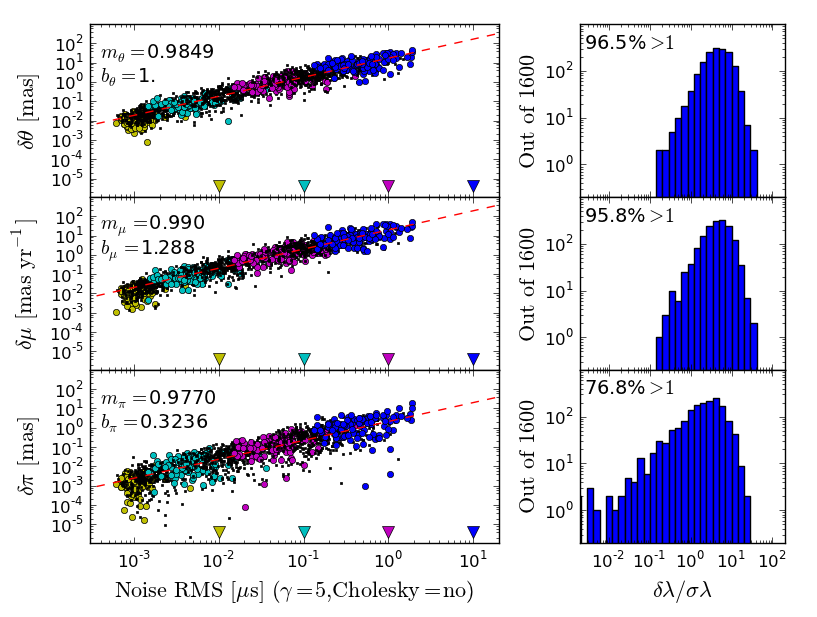}
\includegraphics[height = 74mm,width = 86mm]{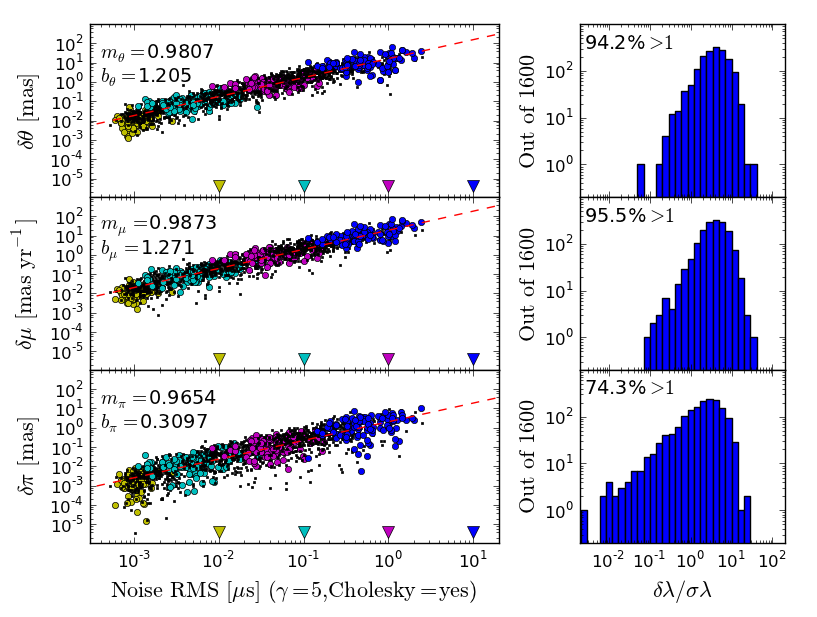}
\caption{As in Figure 3, but for simulation Set B (500 days of TOAs taken at 21 day intervals).}
\end{center}
\end{figure*}

\begin{figure*}[h!tbp]
\begin{center}
\includegraphics[height = 74mm,width = 86mm]{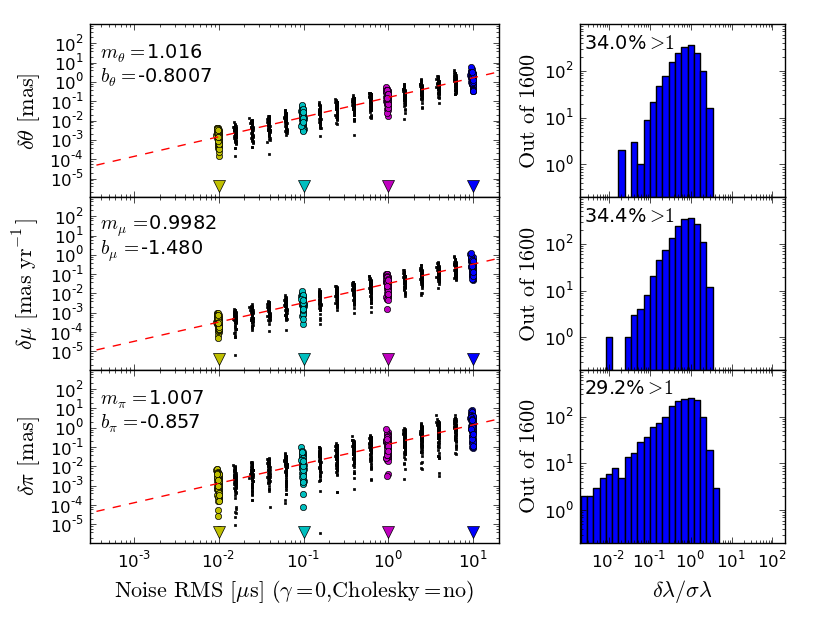}\\
\includegraphics[height = 74mm,width = 86mm]{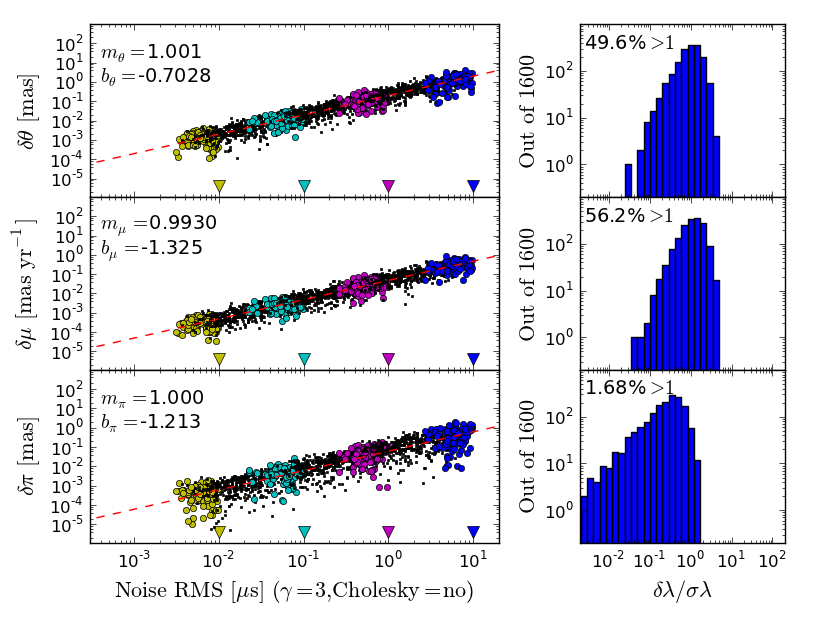}
\includegraphics[height = 74mm,width = 86mm]{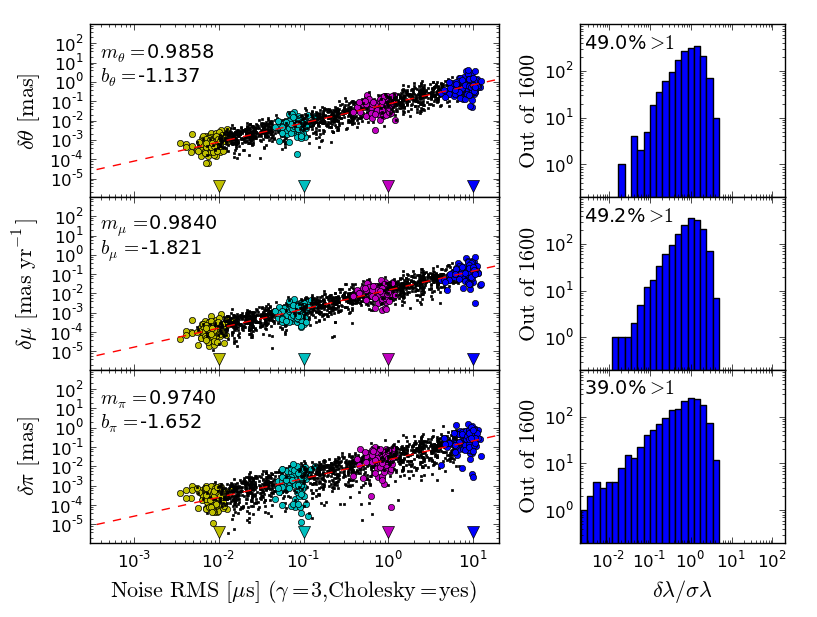}\\
\includegraphics[height = 74mm,width = 86mm]{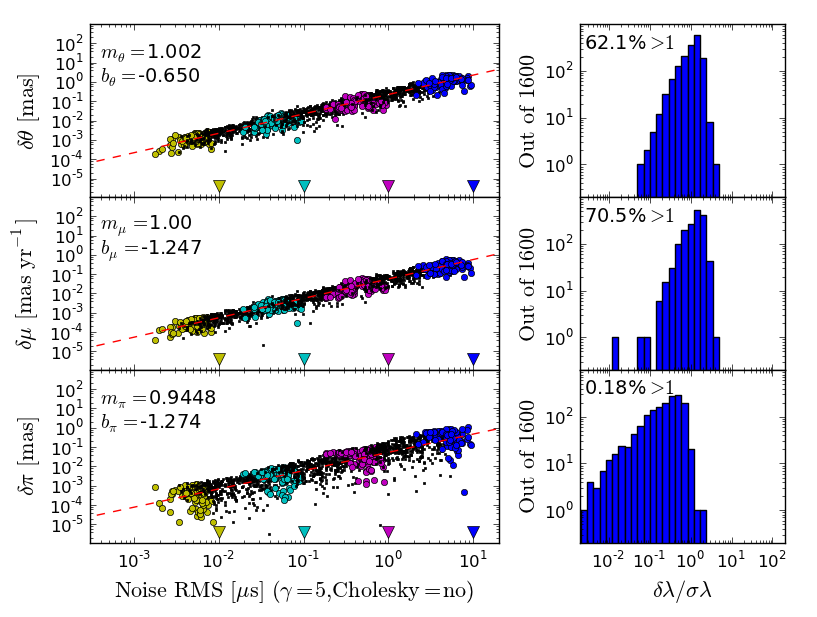}
\includegraphics[height = 74mm,width = 86mm]{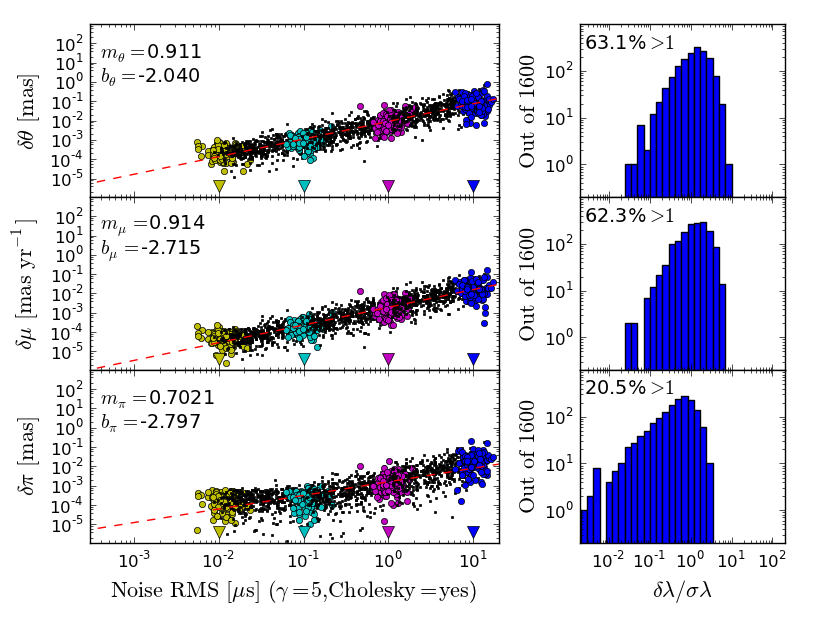}
\caption{As in Figure 3, but for simulation Set C (3000 days of TOAs taken at 21 day intervals).}
\end{center}
\end{figure*}

\begin{deluxetable*}{ccccccccccccc}[h!]
\tablehead{
	\colhead{}&\colhead{}&\colhead{}&\colhead{}&\colhead{}&\colhead{}&\colhead{}&\colhead{}&\colhead{}&\colhead{}&
	\multicolumn{3}{c}{$\sigma_{\cal R}| \rm{Timing~worse~than~VLBI}$}\\\cline{11-13}
	\colhead{}&\colhead{}&\colhead{}&\colhead{}&\colhead{}&\colhead{}&\colhead{}&\colhead{}&\colhead{}&\colhead{}&
	\colhead{$\sigma_{\cal R}|(\delta\theta$}&\colhead{$\sigma_{\cal R}|(\delta\mu$}&\colhead{$\sigma_{\cal R}|(\delta\pi$}\\
	\colhead{NDays}&\colhead{gap}&\colhead{$\gamma$}&\colhead{Cholesky}&\colhead{$m_\theta$}&\colhead{$b_\theta$}&
	\colhead{$m_\mu$}&\colhead{$b_\mu$}&\colhead{$m_\pi$}&\colhead{$b_\pi$}&\colhead{$\sim 1 {\rm ~mas})$}&
	\colhead{$\sim 10~\mu{\rm as~yr}^{-1})$}&\colhead{$\sim 10~\mu{\rm as})$}\\
	\colhead{(days)}&\colhead{(days)}&\colhead{}&\colhead{}&\colhead{}&\colhead{}&\colhead{}&\colhead{}&\colhead{}&
	\colhead{}&\colhead{(ns)}&\colhead{(ns)}&\colhead{(ns)}}
\startdata
500         &3             &5  &no    &   0.967   &   1.18       &   0.970   &   1.25       &   0.991    &    0.363    &    60.2       &    0.448    &    4.13    \\     
500         &3             &5  &yes  &   0.972   &    1.06      &   0.984   &   1.12        &  0.967    &    0.208     &    81.2      &     0.671    &    5.21    \\
500         &3             &3  &no    &   0.982   &   0.859     &  0.991    &   0.939     &  1.00       &    0.166     &   133        &    1.08       &    6.82    \\ 
500         &3             &3  &yes  &   0.975   &    0.664    &   0.982   &   0.765     &   0.996    &   0.0666   &    208        &    1.53      &    8.42    \\
500         &3             &0  &no    &   1.01     &  -0.090    &   1.00      &  -7.05e-3 &   1.01      &   -0.726    &  1.23e3    &     10.2     &    54.8  \\
500         &21          &5  &no    &   0.985   &    1.00      &   0.990   &   1.29        &  0.977    &    0.324     &    96.6        &    0.475   &   4.18    \\
500         &21          &5  &yes  &   0.981   &     1.21     &   0.987   &    1.27       &  0.965     &    0.310    &    58.4        &    0.484   &    4.04    \\
500         &21          &3  &no    &   0.977   &   0.870     &  0.982   &     0.965    &  0.996     &    0.200    &    129         &    0.960   &    6.18   \\
500         &21          &3  &yes  &   0.997   &   0.839     &   0.995   &    0.926    &  0.984      &   0.162     &    144         &    1.15    &    6.35   \\
500         &21          &0  &no    &   0.991   &   0.442     &   0.996   &    0.523    &   0.994    &   -0.188    &    358         &    2.92     &   15.0  \\
3000      &21           &5  &no    &   1.00      &  -0.650    &   1.00     &    -1.25      &   0.945    &   -1.27      &    4.47e3    &   177       &   169  \\
3000      &21           &5  &yes   &   0.911   &  -2.04      &   0.915   &     -2.71     &   0.702    &   -2.80      &   1.73e5     &  6.02e3  &   1.38e4  \\
3000      &21           &3  &no    &    1.00     &  -0.703    &   0.993   &     -1.32     &   1.00      &    -1.21      &   5.05e3    &  209        &   162  \\
3000      &21           &3  &yes   &   0.986   &   -1.14     &   0.984   &      -1.82    &   0.974    &    -1.65     &    1.43e4    &   656       &   437  \\
3000      &21           &0  &no    &    1.02     &   -0.801   &   0.998   &     -1.47     &   1.01      &   -0.857    &    6.10e3   &   294        &   73.8
\enddata
\tablecomments{Best-fit parameters from Figures 3-5 and levels of post-fit noise RMS indicative of astrometric offsets on par with precisions achievable with VLBI.  From left to right, the first two columns indicate the number of days worth of simulated TOAs and the gap (in days) between observations.  The third column indicates the spectral index of the noise power spectrum used for the simulations.  The fourth column indicates whether or not Cholesky whitening was used.  Columns 5-10 record the best-fit scaling parameters from Figures 3-5 (as described in Eq. 8).  The final three columns indicate the post-fit RMS timing residual ($\sigma_{\cal R}$) at which the noise-induced offsets in position, proper motion, and parallax reach approximately 1 mas, 10 $\mu$as yr$^{-1}$, and 10~$\mu$as respectively.  VLBI can achieve position estimates accurate to 1 mas after a single observation; proper motion and parallax can be constrained with VLBI to 10 $\mu$as yr$^{-1}$ and 10 $\mu$as respectively after $\sim$ 8 observation epochs distributed over 2-3 years.  If $\sigma_{\cal R}$ exceeds the values in these last three columns and the time span, cadence, and noise model are similar to what we have simulated, VLBI can measure that astrometric parameter more accurately than timing. The ``$|$'' symbol should be read as ``such that''.}
\end{deluxetable*}

\section{Discussion of Simulation Results}
In this section, we will discuss two main aspects of our simulations: how the magnitude of red processes in timing residuals can be suppressed by basic model fitting and how the astrometric parameters of the model can become biased through such fitting in the presence of noise. In Figure 2, we show histograms of the factors by which the pre-fit RMS amplitudes of the noise are rescaled by fitting. For simulation Sets A and B (top and middle plots of Figure 2), which correspond to 500 days of observations, significant amounts of red noise are absorbed into the model.  In both Set A and Set B, the redder the noise (i.e. the larger the power spectral index), the more severe the absorption is on average.  With the reddest noise, the pre-fit RMS amplitude of the red noise is reduced by a factor of about 20 in the median case.  For these 500-day simulations with both $\gamma = 3$ and $\gamma = 5$ red noise, if we only fit for spin parameters and exclude the astrometric fit, the amount by which the pre-fit RMS is reduced by fitting is approximately halved compared with the case when we simultaneously fit for the spin parameters and the astrometric parameters.  For example, in simulation Set A, if we only fit for phase, period, and period derivative in 100 realizations of $\gamma = 3$ red noise with a pre-fit RMS of 100 ns, the average post-fit RMS is 62 ns (a 38~ns reduction) as compared to an average post-fit RMS of 22~ns (a 78 ns reduction) when we simultaneously fit for the astrometric parameters.  The problem of noise power absorption is less severe in the simulations with longer observation spans (Set C). As the number of expected annual cycles from astrometric errors grows, the signature of an astrometric error becomes more distinct and less covariant with a red noise process.  In these 3000-day simulations, if we instead fit only for the phase, period, and period derivative of the pulsar, the reduction in RMS is effectively the same compared with the case in which we also fit for the astrometric parameters; the power absorbed by the astrometric fitting is negligibly small.

Of the six plots in Figure 2, the three on the right were generated with and the three on the left were generated without Cholesky whitening.  For simulations of 500-day data sets (the top two rows of plots in Figure~2), whether Cholesky whitening is used or not makes only a very small difference to the distribution of post-fit RMS values.  However, a bigger percentage of the RMS reduction is due to the astrometric fit.  In 100 realizations of $\gamma = 3$ noise with a pre-fit RMS of 100 ns, fitting for spin parameters alone reduces the RMS to 79 ns on average as opposed to 26 ns on average when spin and astrometric parameters are fit for, and with $\gamma = 5$ noise, the RMS is reduced to 86 ns on average when spin parameters alone are fit for as opposed to only 7 ns when both spin and astrometric parameters are fit for.  With the 3000-day simulated data sets (the bottom row of plots in Figure~2), Cholesky whitening does noticeably decrease the amount by which red noise power is absorbed into the timing model, and the absorption is more reduced in the case of the $\gamma=5$ noise than in the case of the $\gamma=3$ noise.  In fact, with 3000-day simulations, when Cholesky whitening is used, the post-fit RMS can increase relative to the pre-fit RMS.  With $\gamma=3$ red noise the maximum increase is by a factor of 1.46 and with $\gamma=5$ red noise the maximum increase is by a factor of 2.37.  \citet{chc+11} describe how such increases of the post-fit RMS relative to the pre-fit RMS can occur with Cholesky whitening if the period and period derivative of the pulsar are fit for; such behavior is not surprising. \citet{vl13} describe a Bayesian strategy to fit timing models in the presence of correlated noise that performs comparably to Cholesky whitening but slightly mitigates some of the issues Cholesky whitening encounters when pulse period and period derivative are fit in the presence of correlated timing noise. 

Figures 3, 4, and 5 (corresponding to simulation Sets A, B, and C, respectively) show offsets from the true astrometric parameters in the post-fit astrometric parameter estimates of the timing model when they are fit in the presence of the types of noise we have considered and when Cholesky whitening is or is not used.  The histograms in these plots depict the distribution of these post-fit astrometric offsets divided by the 1-$\sigma$ parameter uncertainty from the fit; in many cases this value is greater than 1, it is greater than 1 more often for redder noise, and it is greater than 1 less often if Cholesky methods are employed. In all cases, the noisier the pre-fit residuals, the noisier the post-fit residuals and the bigger the offsets in the post-fit astrometric parameters.  These figures also depict the reduction of the RMS noise level that is caused by fitting (the horizontal location of the points in these plots indicate the post-fit RMS of the residuals as compared with the pre-fit RMS indicated by the colored markers along the abscissa).  Like in Figure 2, it is apparent that the absorption of red noise is a more severe problem for the 500-day simulations in Sets A and B than it is for the 3000-day simulations of Set C; this can be straightforwardly explained.  

Fitting for phase, period, and period-derivative removes a quadratic from the pre-fit residuals (this simple statement is slightly complicated by Cholesky whitening as the signature of offsets in these parameters is difficult to disentangle from a model for the lowest-frequency noise in the residuals; see \cite{vl13} for a discussion of slightly improved methods in this regard).  The post-fit residuals will thus consist of cubic and higher-order contributions.  The higher order contributions will be suppressed in red processes that have less power in higher frequencies.  A zero-mean polynomial dominated by a cubic contribution can readily mimic a sinusoid with a one year period if the data set has a length of approximately one year.  If the data set is significantly longer than one year, this is not such a problem.  Similarly, if one has only 6 months of data, quadratic subtraction may leave a shape in the residuals resembling the 6-month-period sinusoid associated with errors in parallax.  To quantitatively describe and compare Figures 3, 4, and 5, we have assumed that 
\begin{eqnarray}
\left(\frac{\delta\theta}{1 \rm~mas}\right)&=&10^{b_\theta}\left(\frac{\sigma_{\cal R}}{1~\mu\rm s}\right)^{m_\theta},\nonumber\\
\left(\frac{\delta\mu}{1 \rm~mas~yr^{-1}}\right)&=&10^{b_\mu}\left(\frac{\sigma_{\cal R}}{1~\mu\rm s}\right)^{m_\mu}, \nonumber\\
\left(\frac{\delta\pi}{1 \rm~mas}\right)&=&10^{b_\pi}\left(\frac{\sigma_{\cal R}}{1 \mu\rm s}\right)^{m_\pi},
\end{eqnarray}
where $\sigma_{\cal R}$ is the post-fit RMS of the residuals.  The least-squares estimates of the scaling exponents in Eq. 8 are included as part of Table 1 and these best-fit relations are plotted in Figures 3 through 5 as dashed red lines.  In all cases, $m_\lambda$  (for $\lambda = \theta$, $\mu$, or $\pi$) is very nearly unity.  The more interesting behavior is in the $b_\lambda$ parameters which determine the scale of the astrometric errors for a given post-fit RMS.  When Cholesky whitening is used, $b_\lambda$ is always smaller than when it is not used, but for 500-day data sets, the reduction is not as significant as it is with 3000-day data sets.  As an example, with a post-fit RMS of 1 $\mu$s and $\gamma=5$ noise, the anticipated position error with 500 days of data (and 3-day observing cadence) is approximately 15 mas if Cholesky whitening is not used and approximately 11 mas if it is used; with 3000 days of data, the anticipated position error is approximately 224~$\mu$as if Cholesky whitening is not used and approximately 9 $\mu$as if it is used.

In Figures 3, 4, and 5, there is a stark difference between $\gamma = 0$ (white noise) and the $\gamma = 3$ and $5$ cases.  With white noise, the 100 points associated with each of our 16 levels of input pre-fit RMS noise fall into significantly more compact regions of the plane for position, proper motion, and parallax.  These regions are particularly compact in the horizontal dimension.  All realizations of white noise are comparably non-covariant with the signatures of astrometric errors in the timing model.  On the other hand, some realizations of red noise can be more covariant with the signatures of astrometric errors in the timing model than other realizations.  Some realizations of red noise can be more readily attenuated in post-fit residuals than others by modifying the astrometric parameters of the timing model; this leads to the increased horizontal scatter in simulations involving red noise.
 
\begin{deluxetable*}{cclll}
\tablewidth{0pc}
\tablecolumns{5}
\tablehead{
	\colhead{PSR}&
	\colhead{Measurement Type}&
	\colhead{$\Delta\alpha$}&
	\colhead{$\Delta\delta$}&
	\colhead{Reference}\\
	\colhead{}&
	\colhead{}&
	\colhead{(s)}&
	\colhead{($''$)}&
	\colhead{}}
\startdata
J0437$-$4715	&VLBI             &0.883250(3)	&0.031863(37)	&\citet{dvt+08}\\
. . .			&Timing	      &0.883185(6)	&0.034033(70)	&\citet{vbv+08}\\
			&		      &				&			&                        \\
J1713+0747	&VLBI	      &0.5306(1)		&0.519(2)		&\citet{cbv+09}\\
. . .			&Timing 1	      &0.5307826(7)	&0.52339(3)	&\citet{sns+05}\\
. . .			&Timing 2	      &0.53077(1)	&0.5228(2)	&\citet{hbo06}
\enddata
\tablecomments{Astrometric positions measured with VLBI and pulsar timing for PSR J0437$-$4715 and PSR J1713+0747.  For PSR J0437$-$4715, $\Delta\alpha$ is the offset from right ascension $04^h37^m15^s$ and $\Delta\delta$ is the offset from declination $-47^\circ15'09''$ at MJD epoch 54100. For PSR J1713+0747, $\Delta\alpha$ is the offset from right ascension $17^h13^m49^s$ and $\Delta\delta$ is the offset from declination $07^\circ47'37''$ at MJD epoch 52275.}
\end{deluxetable*}
 
\section{Incorporating VLBI Astrometry Into PTAs}
Given a sufficient S/N and adequately proximal calibrators, modern pulsar-optimized VLBI can provide absolute position estimates in the International Celestial Reference Frame (ICRF) to $\sim$1 mas precision after a single observation; proper motion and parallax can be measured to 10 $\mu$as yr$^{-1}$ and 10 $\mu$as respectively after approximately 8 observing epochs spread over 2-3 years \citep[e.g.][]{cbv+09}.  Chatterjee et al. explain that pulsar gating can typically boost the S/N for pulsars by a factor of three, but particularly faint pulsars must be observed at low frequencies to achieve a sufficiently high flux to do VLBI; at lower frequencies, errors from differential ionospheric activity between the pulsar and the calibrator are the biggest source of systematic error in astrometric measurements.  The PSR$\pi$ project \citep{dbc+11}, an effort with the VLBA to measure precision parallaxes for roughly 200 pulsars, finds that most pulsars in their sample are sufficiently bright for VLBI at 1.6 GHz.  At this frequency, in-beam calibrators are required.  With the relatively wide beam of the VLBA, in 77 hours of observations, they have found one or more satisfactory in-beam calibrators for 97\% of their 200 targets.  While newly discovered pulsars are likely to be dim to have gone undetected in previously conducted surveys, large projects like PSR$\pi$ are increasing the number of pulsars for which precision VLBI-based astrometry has been or can be done by finding calibration sources suitable for lower-frequency observations where pulsars are brighter. Ongoing VLBA sensitivity improvements and the inclusion of the VLBA in the High Sensitivity Array (an effort to combine large, very sensitive telescopes like the Green Bank Telescope, Arecibo, the phased Very Large Array, and Effelsburg into a VLBI-capable array) will allow large samples of pulsars to be addressed.

From Figures 3, 4, and 5, we can infer the post-fit RMS noise amplitudes at which noise-induced errors in the astrometric parameters reach levels comparable to the precisions achievable with VLBI.  A summary of this is presented in Table~1.  The best-timed pulsars to date have post-fit RMS residuals of several tens of nanoseconds over several year spans; several hundred nanoseconds RMS is more typical \citep{ych+11,vlj+11,dfg+13}.  In our 500-day simulations with $\gamma=3$ or 5 red noise, we find that the positions of all but the best-timed pulsars can be more accurately measured with VLBI than with timing.  With 3000 days of data, timing yields more precise positions than VLBI so long as the pulsar has an RMS timing residual less than several microseconds.  With 500 days of data, VLBI provides more precise proper motion measurements (and parallax unless the noise is white and the pulsar has less than 50 ns RMS timing residuals).  With 3000 days of data, if Cholesky whitening is used, timing generally produces more precise proper motion and parallax measurements, but not for pulsars displaying several hundred nanoseconds RMS of $\gamma=3$ red noise or white noise.  With these observations, it is apparent that VLBI can play a significant role in better establishing the astrometric parameters of many pulsars that are currently in PTAs (especially proper motion and parallax), and it can be especially important for relatively newly discovered pulsars with short observation baselines.

\subsection{Reference Frame Considerations}
Astrometric measurements with VLBI are made relative to the ICRF \citep{mae+98}, a geocentric, quasi-inertial frame referenced to distant quasars.  Pulsar timing is carried out in a dynamic reference frame centered on the SSB and oriented relative to the orbital plane of Earth; this coordinate system is set by a solar system ephemeris such as the Jet Propulsion Laboratory's DE405 \citep{s98}.  Position measurements of some pulsars with VLBI and pulsar timing are inconsistent, indicating that the quoted measurement errors are too small.  To demonstrate the inconsistency between VLBI and timing-based astrometry, in Table~2, we show VLBI and timing-based position measurements of pulsars J0437$-$4715 and J1713+0747.  In both cases, timing-based estimates and VLBI estimates become inconsistent at the milliarcsecond level while quoted errors are typically a few tens of microarcseconds.  The inconsistency is less severe for J1713+0747 primarily due to the relatively large error estimate of Chatterjee et al.   The two timing-based position estimates for pulsar J1713+0747 in Table~2 are also inconsistent with one another at the milliarcsecond scale (a considerably larger scale than the quoted errors). This inconsistency is also due, at least in part, to reference frame differences as \citet{sns+05} use the DE200 \citep{s90} solar system ephemeris while \citet{hbo06} use the more recent DE405 ephemeris.  Proper motion and parallax measurements are less susceptible to such frame-related offsets as they depend more on the changes in celestial coordinates over time than the absolute coordinates.  However, as Eq. 4 demonstrates, timing corrections associated with proper motion and parallax depend on the position of the pulsar, so reference frame mismatch will influence the degree to which VLBI-based and timing-based proper motion and parallax measurements can be made to agree.

By measuring the absolute position of several pulsars with both timing and VLBI at approximately contemporaneous epochs, it is possible to constrain the three free parameters defining the frame tie between positions in the ICRF and positions in the relevant pulsar timing frame to milliarcsecond precision or better \citep[e.g.,][]{bcr+96}.  The frame tie is an orthogonal matrix ${\bf \Omega}$ that transforms positions measured in the ICRF to positions in the relevant pulsar timing frame.  If the two frames differ by only very small rotations, ${\bf \Omega}$ can be approximated as ${\bf \Omega}\approx{\bf I}+{\bf \Lambda}$ where 
\begin{eqnarray}
{\bf \Lambda}=\left(\begin{matrix} 0 & A_z & -A_y \\ -A_z & 0 & A_x \\ A_y & -A_x & 0 \end{matrix}\right)
\end{eqnarray}
and $A_x, A_y,$ and $A_z  \ll 1$. To constrain these parameters, consider a VLBI position measurement ${\bf\hat{n}_i}$ with uncertainty ${\bf\epsilon_i}$ and a timing position measurement ${\bf\hat{n}_i'}$ with uncertainty ${\bf\epsilon_i'}$.  The index $i$ indicates the $i^{\rm th}$ pulsar.  We want ${\bf \Omega}$ such that
\begin{eqnarray}
{\bf\hat{n}_i'}+{\bf\epsilon_i'}={\bf \Omega}({\bf\hat{n}_i}+{\bf\epsilon_i})\approx{\bf\hat{n}_i}+{\bf\epsilon_i}+{\bf\Lambda\hat{n}_i}.
\end{eqnarray}
The above equation is an approximation only because we have discarded a second order term proportional to the product of the position uncertainty and the frame tie rotations.  After defining ${\bf d_i}\equiv{\bf\hat{n}_i'}-{\bf\hat{n}_i}$ and ${\bf e_i}\equiv{\bf\epsilon_i}-{\bf\epsilon_i'}$, Eq. 10 can be written as ${\bf d_i}={\bf\Lambda\hat{n}_i}+{\bf e_i}$ which can in turn be written as ${\bf d_i}={\bf m_i\tilde{A}}+{\bf e_i}$ where ${\bf\tilde{A}^T}=[A_x~A_y~A_z]$ and 
\begin{eqnarray}
{\bf m_i}=\left(\begin{matrix} 0 & -\hat{{\rm n}}_{i,z} & \hat{{\rm n}}_{i,y} \\ \hat{{\rm n}}_{i,z} & 0 & -\hat{{\rm n}}_{i,x} \\ -\hat{{\rm n}}_{i,y} & \hat{{\rm n}}_{i,x} & 0 \end{matrix}\right).
\end{eqnarray}
With position measurements of only one pulsar, the frame tie can not be constrained because the determinant of ${\bf m_i}$ is zero, but if $N_P\geq 2$, we can define ${\bf D^T}\equiv[{\bf d_1^T}\cdots{\bf d_{N_P}^T}]$, ${\bf E^T}\equiv[{\bf e_1^T}\cdots{\bf e_{N_P}^T}]$, ${\bf\Sigma}~\equiv~\langle{\bf E^TE}\rangle$, and ${\bf M^T}\equiv[{\bf m_1^T}\cdots{\bf m_{N_P}^T}]$ to compute
\begin{eqnarray}
{\bf\hat{A}}=\left({\bf M^T\Sigma}^{-1}{\bf M}\right)^{-1}{\bf M^T\Sigma}^{-1}{\bf D},
\end{eqnarray}
the best least-squares estimate for the frame tie parameters.  

With the Deller et al., Verbiest et al., Chatterjee et al., and Hotan et al. measurements in Table 2, we have used the above formalism to initially constrain the rotation parameters: $A_x\approx2.26$ mas, $A_y\approx-0.36$ mas, and $A_z\approx0.53$ mas.  We have approximated ${\bf \Sigma}$ as diagonal; since VLBI and timing measurement uncertainties are uncorrelated and measurement uncertainties between pulsars using either VLBI or timing are uncorrelated, this assumption only means that we have ignored correlations in the VLBI-based or timing-based measurement uncertainties of right ascension and declination for individual pulsars.  With these numbers, ${\bf \Omega}$ transforms the VLBI position of J0437$-$4715 to within 0.1 mas of the timing based position (as opposed to the initial, untransformed offset of 2.26 mas) and it transforms the VLBI position of J1713+0747 to within 3.4 mas of the timing based position (as opposed to the initial, untransformed offset of 4.56 mas).  The smaller positional errors for J0437$-$4715 means our weighted transformation responds more to the position offsets for J0437$-$4715 than to the position offsets for J1713+0747. With more precise VLBI measurements of J1713+0747, the transformed position offset could be constrained to approximately 100 microarcseconds like that of J0437$-$4715.  

A higher precision transformation is needed if VLBI measurements are to be made useful in pulsar timing models.  If comparably precise VLBI and timing astrometry is measured for $N_P$ pulsars and the measurement uncertainties are uncorrelated from pulsar to pulsar, the uncertainties in the frame transformation can be reduced by a factor $\propto N_P^{-1/2}$. With large projects like PSR$\pi$ \citep{dbc+11} currently measuring proper motions and parallaxes for hundreds of pulsars, the precision with which the frame-tie transformation can be constrained will soon approach the microarcsecond precisions necessary to incorporate VLBI astrometry into pulsar timing models.  At the outset of a timing campaign on a sample of newly discovered pulsars, pre-existing measurements such as those presented in Table~2 and those that will come out of efforts like PSR$\pi$ can be used to refine the frame tie to an extent that would allow VLBI astrometry to immediately be used in timing models in place of estimates derived from timing data.  As TOAs are collected, timing models should be refined using standard techniques, but the astrometric parameters of the model will not have to be fit for.  We will elaborate upon this in future work as more data becomes available.

Once the time baseline has significantly exceeded one year, the PTA data can be made independent of the pre-existing frame tie by combining the timing data and the VLBI astrometric measurements and doing one simultaneous, global fit over all pulsars and the frame tie.  Using timing residuals from all pulsars in a PTA to constrain the three frame-tie parameters will require incorporating currently existing tools like TEMPO2 into a more elaborate framework that can simultaneously fit the timing models of all the pulsars in the array as well as the frame-tie parameters with appropriate weighting to account for different uncertainties.  For a PTA with 17 pulsars, \citet{dfg+13} fitted 1382 parameters.  Of these parameters, 1171 dealt with a time-varying dispersion measure or pulse profile evolution across the band.  With more thorough multi-frequency observations, many of these parameters will become unnecessary.  The remaining frequency-dependent parameters will be determined by features unique to the residuals of individual pulsars and will prove highly non-covariant with parameters that affect the quality of the global fit.  Of the remaining 211 parameters, 85 are astrometric and could be provided by VLBI if three additional frame-tie parameters were included in the fitting procedure.

\subsection{Incorporating New Pulsars Into PTAs While Utilizing VLBI Astrometry}
Suppose a new MSP is discovered and its timing stability and brightness make it a prime candidate for PTA science.  In the first couple of years of observations, if timing and VLBI studies begin concurrently and are combined appropriately, attenuation of the amplitude of red processes in the timing residuals caused by model fitting can be dramatically reduced.  To quantify this, we introduce some formalism taken from \citet{brn84}.  The timing residuals, $R(t)$, are fit to a linear combination of basis functions associated with the parameters being fit for.  To simplify matters, consider only eight parameters as follows:
\begin{equation}
R(t)=\sum_{i=1}^8 \lambda_i\psi_i(t);
\end{equation}
\begin{equation}
\begin{aligned}[c]
\psi_1(t)&=1\nonumber\\
\psi_2(t)&=t\nonumber\\
\psi_3(t)&=t^2\nonumber\\
\psi_4(t)&=\sin{(\omega t)}\nonumber\\
\end{aligned}
~~~~~
\begin{aligned}
\psi_5(t)&=&\cos{(\omega t)}\nonumber\\
\psi_6(t)&=&t \cos{(\omega t)}\nonumber\\
\psi_7(t)&=&t \sin{(\omega t)}\nonumber\\
\psi_8(t)&=&\sin{(2\omega t)}.\\
\end{aligned}
\end{equation}
The basis functions $\psi_i(t)$ are associated with $\phi,P,\dot{P},\alpha,\delta,\mu_\alpha,\mu_\delta,$ and $\pi$ respectively.  The coefficients $\lambda_i$ are functions of the true spin and astrometric parameters of the pulsar and the fractional errors in the initial timing model estimates of them.  Assuming $N$ years of observations have been made, the basis functions can be converted to an orthonormal set of basis functions, $\psi'_i(t)$, using the Gram-Schmidt process with the inner product
\begin{eqnarray}
\psi_i(t)\cdot\psi_j(t)=\int_{-N/2}^{N/2}\psi_i(t)\psi_j(t)dt.
\end{eqnarray}
We then compute the Fourier transform of each basis function, $\tilde{\psi}_i'(f)$, and combine them into a transmission function, $T(f)$ as follows: 
\begin{eqnarray}
\tilde{\psi}_i'(f)&=&\int_{-N/2}^{N/2}\psi_i'(t)\exp{(2\pi \imath ft)}dt,\\
T(f)&=&1-\frac{1}{N}\sum_{i=1}^8\tilde{\psi}_i'(f)\tilde{\psi}_i'^*(f).
\end{eqnarray}
If the structure in the pre-fit timing residuals is due entirely to some process with a stationary power spectrum $P(f)$, the mean-square post-fit residual coming from frequencies between $f_L$ and $f_H$ is given by
\begin{eqnarray}
\bar{R^2}&=&\int_{f_L}^{f_H} T(f)P(f)df.
\end{eqnarray}
The transmission function is a filter on the pre-fit variance and dictates what fraction of power in the pre-fit residuals at a particular frequency makes it through the fitting procedure and into the post-fit residuals; it does not, however, indicate the magnitude of the errors induced in the timing model parameters from the power being absorbed into the timing model.

\begin{figure}[tbp]
\begin{center}
\includegraphics[height = 75mm,width = 90mm]{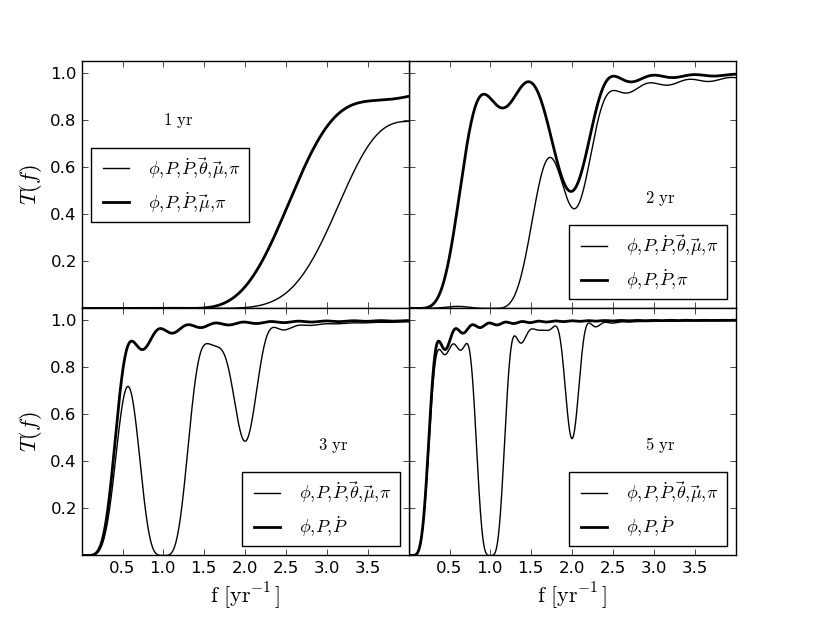}
\caption{Transmission functions for 1, 2, 3, and 5 year data sets with and without using astrometric measurements from VLBI. The legends indicate the spin and astrometric parameters being fit to produce the associated transmission function.  In the 5 year transmission function in which all 8 parameters are fit, the prominent absorption feature at the origin ($f = 0$ yr$^{-1}$) is associated with the quadratic fit to the spin parameters of the pulsar, the feature at 1 yr$^{-1}$ is associated with fitting out the oscillations associated with errors in position and proper motion, and the feature at 2 yr$^{-1}$ is associated with fitting out parallax errors.}
\end{center}
\end{figure}

\begin{figure}[tbp]
\begin{center}
\includegraphics[height = 75mm,width = 90mm]{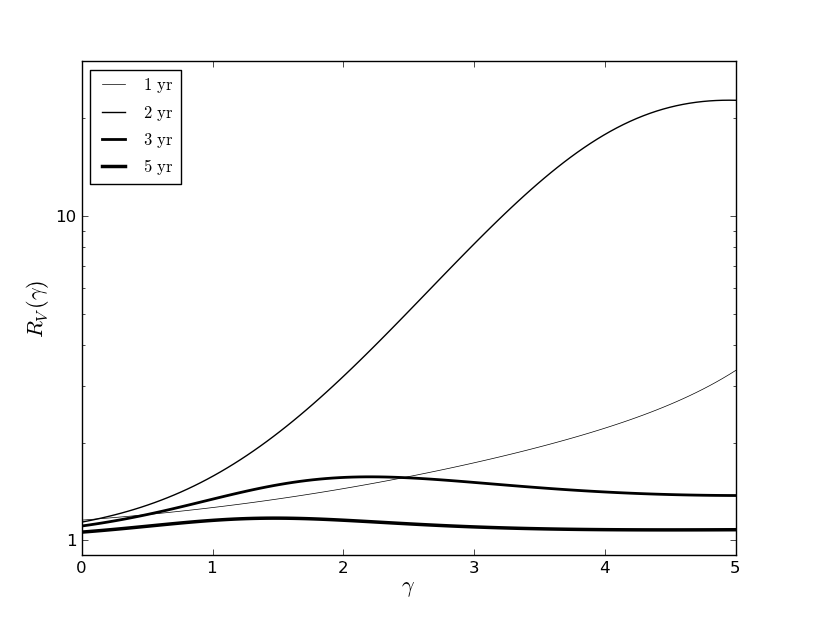}
\caption{The ratio of post-fit residual variance using VLBI astrometry to that without using VLBI astrometry versus the spectral index of a red process in the TOAs.  Each curve is associated with one panel from Figure 6, as is indicated by the legend. It is assumed that astrometric parameters from VLBI are incorporated into the timing model according to the same schedule used to make Figure 6; after 1 year, VLBI provides $\boldsymbol{\theta}$, after 2 years, it also provides $\boldsymbol{\mu}$, and at the 3 and 5 year marks, VLBI is providing all 5 astrometric parameters.  The quantity being plotted is $R_V$ from Eq. 18.}
\end{center}
\end{figure}

If sufficiently precise VLBI astrometry is available and reference frame mismatches have been dealt with, the astrometric parameters of the timing model do not need to be fit for.  The process described above can be repeated with any subset of the basis functions listed in Eq. 13.  We assume that after the first year of timing and VLBI observations, VLBI estimates for position are precise enough that they do not need to be fit for. After 2 years, the VLBI estimates for proper motion reach sufficient precision, and after 3 years, VLBI can determine parallax well enough that no astrometric fitting is necessary.  In Figure 6, we demonstrate how the transmission function varies in the first 5 years of observation if VLBI astrometry is or is not incorporated into timing efforts according to this schedule. 

The sensitivity enhancements when VLBI astrometry is used are apparent in Figure 6.  Because the bulk of the sensitivity enhancement occurs at low frequencies (below about 3~yr$^{-1}$), the sensitivity of PTAs to red processes can be greatly enhanced.  To quantify the enhancement to red process sensitivity, for a range of spectral indices $\gamma$, we compute the following variance ratio:
\begin{eqnarray}
R_V(\gamma)&=&\frac{\int_{f_L}^{f_H}T_{VLBI}(f)f^{-\gamma}df}{\int_{f_L}^{f_H}T(f)f^{-\gamma}df}
\end{eqnarray}
where $T_{VLBI}(f)$ is the transmission function when VLBI astrometry is used and $T(f)$ is the transmission function when all the astrometry is fit for.  We set the low frequency cutoff $f_L = 0$.  We set the high frequency cutoff $f_H = 9$ yr$^{-1}$ corresponding to the Nyquist frequency if 18 equispaced measurements are made each year.  The low-frequency absorption feature common to all the transmission functions in Figure 6 approaches zero as $f^6$ for small $f$, so each of the integrals in Eq. 18 will converge for $\gamma < 7$.    The results of this calculation are shown in Figure 7.  As $\gamma$ increases (steeper power spectrum), the increased amplitude of the transmission function at low frequencies when VLBI information is used becomes increasingly important so as to not filter out substantial amounts of the power through fitting.  The enhanced sensitivity is most pronounced after two years with red processes having a spectral index very near 13/3, the value expected from a SB of GWs.  The enhancement can be greater than a factor of 20 in this case.  This is consistent with our findings from Figure 2 that for data sets 500 days in length, red noise with a spectral index of 5 is attenuated by fitting by a factor of 20 in the median case.  

Once the data set becomes longer than three years, the transmission functions with and without VLBI astrometry differ only by the prominent absorption features at 1 and 2 yr$^{-1}$. The width of these features $\Delta f$ is approximately proportional to $f / N$ so the feature at $f = 1$~yr$^{-1}$ has a width $\Delta f\approx N^{-1}$ yr$^{-1}$ and the feature at $f=2$~yr$^{-1}$ has a width $\Delta f\approx 2N^{-1}$ yr$^{-1}$.  The power filtered out by these features diminishes over time and the transmission functions with or without VLBI become essentially indistinguishable.

\section{Additional VLBI and Timing Synergies}
So far, we have emphasized the significance VLBI astrometry can have in PTA efforts to detect GWs.  Here, we briefly discuss additional types of pulsar timing science that could be aided by VLBI astrometry.

\subsection{Relativistic Evolution of NS-NS Binaries}

Pulsars in compact binaries show post-Keplerian (PK) evolution described by parameters that include the rate of periastron advance, the gravitational redshift, and the rate of orbital period decay.  Within any theory of gravity, these PK parameters are functions of the unknown masses of the two objects in the system and the measurable Keplerian parameters.  Measurements of any two independent PK parameters yield estimates of the two masses.  Measurements of additional PK  parameters provide tests of that theory of gravity \citep{s03}.  Studies of the binary systems PSR~B1913+16 \citep{ht75, wnt10}  and PSRs~J0737$-$3039A$/$B \citep{ksm+06} have in this way provided stringent tests of general relativity (GR) and, through measurements of the rate of orbital period decay, $\dot{P}_b$, the best indirect evidence for gravitational radiation to date.  

Measurements of $\dot{P}_b$ are complicated by the fact that, in the timing model, a decaying orbital period is covariant with the relative acceleration along the line of sight between the centers of mass of the compact binary and the SSB.  Additionally, the Shklovskii effect is an apparent radial relative acceleration caused by a changing line of sight to the pulsar that depends on the proper motion and distance to the pulsar as $\mu^2D$ \citep{s69}.  For some particularly close binaries with relatively large velocities transverse to the line of sight, like PSR~B1534+12, the Shklovskii effect can be the dominant bias preventing an accurate measurement of $\dot{P}_b$ \citep{s03}.  If the proper motion and parallax of a pulsar are provided by VLBI, the bias from the Shklovskii effect can be removed to facilitate the study of the relativistic evolution of these systems \citep{dbt09}.

\subsection{Multi-frequency Studies of Newly \\Discovered $\gamma$-ray MSPs}

Recent years have seen an explosion in the population of known MSPs.  This is due, in no small part, to the Large Area Telescope (LAT) on the Fermi Gamma-ray Space Telescope.  \citet{rap+12} describe radio follow-ups of a number of LAT pulsar candidates.  Of 300 candidates that were $\gamma$-ray sources having spectra characteristic of pulsars and signs of variability but no definitive source classification, 47 new pulsars were discovered, 43 of which were MSPs.  In one case, by conducting a radio follow-up with the Nan\c{c}ay Radio Telescope on an unidentified LAT $\gamma$-ray source, \citet{gfc+12} discovered the MSP~J2043+1711.  Using an ephemeris based on timing data from Arecibo, Nan\c{c}ay, and Westerbork, the $\gamma$-ray data were folded and pulsations were detected in the $\gamma$-ray band.  Additional follow-up observations with Suzaku and the Swift X-ray Telescope failed to identify any X-ray counterpart to J2043+1711.  Similarly, \citet{brr+13} discovered PSR J1544+4937 by following up a LAT point source with the Giant Metrewave Radio Telescope (GMRT).  Radio observations allowed the development of a timing model with which $\gamma$-ray counts could be folded and the source was found to be pulsating in $\gamma$-rays.  Follow-up observations in X-rays and optical have failed to detect a counterpart.

Given the rapid rate of discovery of MSPs by instruments like LAT, timing observations are limited by the availability of telescope time.  VLBI astrometry will accelerate and improve multi-wavelength studies of objects like J2043+1711 and J1544+4937.  The $\gamma$-ray sources from LAT are localized to about $1^{\circ}$.  If a telescope like the Green Bank Telescope can detect a radio pulsar counterpart to a $\gamma$-ray source, the localization will improve to a few arcminutes.  At this point, standard pulsar timing techniques can be used to develop a timing model that will allow for, e.g., the detection of pulsations in $\gamma$-rays.  But, until several years of timing data are collected, timing will be unable to constrain the position of the pulsar sufficiently to allow successful identification of a unique X-ray counterpart. Furthermore, there is a chance that a newly discovered pulsar may exhibit high levels of timing noise that substantially skews timing-based astrometric estimates.  After the pulsar is localized to within a few arcminutes by an initial radio detection, we can instead use interferometry (e.g. with the Very Large Array and then VLBI) to constrain the position to milliarcsecond precisions.  Efforts to detect the pulsar in X-rays can be better guided and unique X-ray counterparts can be picked out.  Rapid, high-precision astrometry from VLBI will facilitate efforts like that of \citet{gfc+12} and \citet{brr+13} and will aid studies of the pulsar across the entire electromagnetic spectrum.

\section{Conclusions}
Standard model fitting procedures used in pulsar timing can falter in the presence of red processes.  Whether from red spin noise or the activity of a SB of GWs, fitting the timing model by minimizing the RMS of the residuals causes some power in the red process to be absorbed into the timing model at the cost of shifting the timing model parameters to erroneous values.  Cholesky whitening techniques that account for correlated noise can reduce how much power is absorbed into the timing model, lessening parameter estimation bias and leading to more realistic and often larger parameter error estimates.  However, Cholesky whitening does not radically improve the estimation of astrometric parameters until several years of observations have been accumulated and the signatures of astrometric errors in the timing model become sufficiently non-covariant with red noise-like structure in the residuals.

VLBI provides an alternative means of measuring the astrometric parameters of a pulsar with sufficiently high precision that these measurements can be adopted by timing models without fitting.  A transformation between solar system-based coordinate systems and the ICRF must be derived for this to be feasible.  With the number of pulsars for which precision VLBI-derived astrometric parameters are measured growing quickly because of efforts like PSR$\pi$, soon this transformation will be constrained to within a few microarcseconds.  This timing-independent pulsar astrometry will, among other things, aid efforts to model the Shklovskii effect in some relativistic binaries and thereby allow study of PK models of gravity.  It will also facilitate multi-wavelength studies of the rapidly growing population of MSPs by  providing milliarcsecond sky localization on very short timescales.  

We find that for even the best-timed pulsars to date, if standard model fitting (with or without Cholesky whitening) is employed on data sets not significantly longer than 1~year, the noise-induced errors in the astrometric parameters can exceed the precisions achievable with VLBI in similar time spans.  If VLBI measurements are incorporated into PTA timing models, the reduction in the number of parameters being fit for is a permanent improvement to $\chi^2$ tests for the presence of signal in the data, but this becomes less significant as more TOAs are collected and the number of degrees of freedom in the data grows.  More significantly, by incorporating VLBI astrometry into timing models for newly discovered pulsars, after collecting data for just two years, the power present in post-fit residuals from red processes having spectral indices characteristic of a SB of GWs can be enhanced by a factor of 20.  The ability to augment the power of a SB of GWs in the timing residuals of a pulsar in certain cases by a factor as substantial as 20 may play a key role in the initial direct detection by PTAs of GWs.\\   

\acknowledgements{
We thank A.~Deller, G.~Hobbs, M.~McLaughlin, S.~Ransom, J.~Lazio, and other members of the NANOGrav and PPTA collaborations for much helpful discussion.  We thank P. Demorest and, again, G. Hobbs for answering many questions regarding TEMPO2.  We thank our referee for thoughtful comments and thorough review.  This work was supported by a sub-award to Cornell University from West Virginia University from NSF/PIRE Grant 0968296.  S. Chatterjee also acknowledges support from the NASA Fermi mission through the grant NNX11AO34G.
}

\bibliography{astrometryDIII}
\end{document}